\documentclass[11pt]{article}
\usepackage[utf8]{inputenc}
\usepackage{amsmath,amssymb,amsfonts}
\usepackage{graphicx}
\usepackage{authblk}
\usepackage{hyperref}
\usepackage{cite}
\usepackage{geometry}
\title{A Multi-Agent AI Framework for Immersive Audiobook Production through Spatial Audio and Neural Narration}

\author[1]{Arul Selvamani Shaja, PhD}
\author[2]{Nia D'Souza Ganapathy} 
\affil[1]{Cisco Systems, Inc. San Jose, USA}
\affil[2]{Bombay International School, Mumbai}

\date{\today}

\begin{document}

\maketitle

\begin{abstract}
This research introduces an innovative AI-driven multi-agent framework specifically designed for creating immersive audiobooks. Leveraging neural text-to-speech synthesis with FastSpeech 2 and VALL-E for expressive narration and character-specific voices, the framework employs advanced language models to automatically interpret textual narratives and generate realistic spatial audio effects. These sound effects are dynamically synchronized with the storyline through sophisticated temporal integration methods, including Dynamic Time Warping (DTW) and recurrent neural networks (RNNs). Diffusion-based generative models combined with higher-order ambisonics (HOA) and scattering delay networks (SDN) enable highly realistic 3D soundscapes, substantially enhancing listener immersion and narrative realism. This technology significantly advances audiobook applications, providing richer experiences for educational content, storytelling platforms, and accessibility solutions for visually impaired audiences. Future work will address personalization, ethical management of synthesized voices, and integration with multi-sensory platforms.

\end{abstract}

\section{Introduction}
The integration of artificial intelligence (AI) into multimedia content creation has dramatically transformed audiobook production, enabling unprecedented immersive auditory experiences. This paper presents an advanced AI-based audiobook generation system designed explicitly for immersive storytelling, integrating spatial audio effects that dynamically respond to narrative context. A multi-agent AI approach effectively automates the production pipeline, enhancing listener engagement and realism.

The system leverages specialized agents, including a neural text-to-speech (TTS) agent that uses FastSpeech 2 and VALL-E to produce expressive narration tailored to distinct characters. Concurrently, a spatial narration agent employs sophisticated natural language processing (NLP) to interpret textual cues for generating spatially accurate sound effects. A dedicated spatial audio agent synthesizes these effects using diffusion-based generative models, ensuring alignment and synchronization with the narrated content. Lastly, an audio mixing agent integrates these components into a cohesive immersive audiobook optimized for diverse playback environments.

This research directly addresses the limitations of traditional audiobook production, offering scalable solutions and significantly elevating the listener’s auditory experience.

\section{Background and Related Work}
The evolution of audiobook production has been shaped by advancements in speech synthesis, spatial audio rendering, and AI-driven content generation. This section synthesizes foundational research, recent breakthroughs, and open challenges in the field.

\subsection{Neural Text-to-Speech Synthesis}
Modern TTS systems leverage deep learning architectures to achieve human-like speech quality. The introduction of non-autoregressive models like FastSpeech 2 \cite{ren2020} revolutionized inference speed by eliminating sequential token generation, enabling real-time audiobook production. Subsequent work on zero-shot voice cloning, exemplified by VALL-E \cite{zhang2022}, demonstrated the ability to synthesize personalized narration using only 3-second voice samples, addressing the need for diverse vocal styles in audiobooks. Large-scale implementations, such as Hamilton et al.'s system \cite{hamilton2023}, automated the conversion of 50,000+ e-books into audiobooks while preserving prosody and emotional consistency across chapters. These systems employ self-supervised learning on massive speech datasets (e.g., LibriLight \cite{librilight2019}) to generalize across languages and accents.

\subsection{Spatial Audio Generation Techniques}
Spatial audio research has transitioned from traditional head-related transfer function (HRTF) models \cite{blauert2013} to neural acoustic fields \cite{luo2022} that learn continuous sound propagation functions. Recent work by Huang et al. \cite{huang2023} introduced diffusion-based text-to-audio models capable of generating 3D soundscapes from textual prompts, enabling precise placement of sound sources (e.g., footsteps behind the listener). Temporal synchronization remains critical, with Liu et al. \cite{liu2023} proposing cross-modal attention mechanisms to align sound effects with narrative pacing. These techniques enable dynamic adjustments based on scene descriptions, such as simulating Doppler effects for moving vehicles or reverberation in enclosed spaces.

\subsection{Multi-Agent Architectures for Audio Production}
Modular AI systems have emerged as a paradigm for complex audiobook pipelines. Wang et al. \cite{wang2023} demonstrated how task decomposition into specialized agents (text parsing, voice synthesis, quality control) reduces error propagation. The Speechki framework \cite{abramov2022} combined human editors with AI agents for hybrid quality assurance, achieving 98\% accuracy in defect detection. However, Kim et al. \cite{kim2023} identified emotional consistency as a key challenge, proposing multi-agent reinforcement learning to maintain narrative tone across chapters. These systems often employ distributed computing frameworks like Ray \cite{moritz2018} to parallelize tasks such as phoneme alignment and prosody prediction.

\subsection{Ethical and Market Considerations}
The rise of AI-narrated audiobooks has disrupted traditional production models. Chen et al. \cite{chen2023} quantified a 6.3\% annual decline in human-narrated content consumption, correlating with advancements in neural TTS quality. Ethical concerns around voice cloning have prompted solutions like Li et al.'s \cite{li2023} watermarking system, which embeds inaudible identifiers in AI-generated audio. Smith et al. \cite{smith2023} highlighted legal risks in unauthorized voice replication, advocating for blockchain-based consent management. Industry analyses project a \$5.2B market for AI-narrated audiobooks by 2027 \cite{hiig2025}, driven by demand for multilingual and accessible content.

\section{Study of Existing Approaches in Audiobook Generation, Spatial Audio Synthesis, and Multi-Agent Systems}
The rapid advancements in artificial intelligence (AI) have revolutionized audiobook production, spatial audio synthesis, and multi-agent systems, enabling more efficient, scalable, and immersive content creation. This section discusses the current state of these technologies, highlighting their strengths, limitations, and areas for improvement.

\subsection{Audiobook Generation Approaches}
Traditional audiobook production has been a labor-intensive process involving professional narrators, recording studios, and extensive post-production editing \cite{loc2025}. While this approach ensures high-quality narration with emotional depth and human-like delivery, it is costly and time-consuming. The emergence of AI-driven text-to-speech (TTS) systems has transformed audiobook generation by automating the narration process.

Modern TTS systems leverage neural architectures to produce natural-sounding voices with expressive intonation. Tools such as Murf AI, ElevenLabs, and PlayHT offer customizable options for pitch control, tone modulation, and emotional expression \cite{murf2025}. Neural TTS models like FastSpeech 2 \cite{ren2020} and VALL-E \cite{zhang2022} have further enhanced the quality of synthesized speech by introducing non-autoregressive generation and zero-shot voice cloning capabilities. These systems allow for rapid audiobook production while maintaining high perceptual quality.

Despite these advancements, challenges remain. AI-generated voices often lack the emotional nuance of human narrators, particularly in complex narratives requiring dynamic vocal performance \cite{marktechpost2023}. Additionally, ethical concerns around voice cloning and listener preferences for human-like narration continue to shape the adoption of AI in audiobook production \cite{smith2023}.

\subsection{Spatial Audio Synthesis Techniques}
Spatial audio synthesis has become a cornerstone of immersive storytelling by simulating three-dimensional soundscapes that enhance listener engagement. Traditional methods relied on head-related transfer functions (HRTFs) to model sound localization cues \cite{blauert2013}. However, these approaches were computationally expensive and limited in their ability to adapt to dynamic environments.

Recent advancements in neural spatial audio synthesis have addressed these limitations. Generative models like Make-An-Audio \cite{huang2023} utilize diffusion-based techniques to create realistic soundscapes from textual descriptions. Temporal-enhanced text-to-audio generation frameworks synchronize sound effects with narrative pacing using cross-modal attention mechanisms \cite{liu2023}. These innovations enable precise placement of sound sources and dynamic adjustments based on scene context.

Despite these advancements, spatial audio synthesis faces challenges in achieving real-time performance and adapting to listener-specific environments (e.g., headphones vs. speakers). Furthermore, the integration of spatial audio into audiobooks requires seamless alignment with narrative content to avoid dissonance between narration and sound effects \cite{luo2022}.

\subsection{Multi-Agent Systems for Audiobook Production}
Multi-agent systems have emerged as a modular solution for complex audiobook production pipelines. By dividing tasks among specialized agents, these frameworks optimize efficiency and scalability. For instance:
\begin{itemize}
    \item A text-to-speech agent converts written text into spoken audio using advanced TTS algorithms.
    \item A natural language processing (NLP) agent interprets narrative cues to generate descriptions of spatial sound effects.
    \item A generative audio agent synthesizes spatial audio based on these descriptions.
    \item A mixing agent combines the synthesized audio with narrated text to produce a cohesive output.
\end{itemize}

Wang et al. \cite{wang2023} demonstrated the effectiveness of multi-agent systems in reducing error propagation and improving task parallelization. Speechki's hybrid human-AI framework further highlighted how multi-agent architectures can enhance quality assurance through collaborative workflows \cite{abramov2022}. However, maintaining emotional consistency across agents remains a significant challenge, as identified by Kim et al. \cite{kim2023}, who proposed reinforcement learning-based solutions to address this issue.

\subsection{Human-in-the-Loop in Audiobook}
The integration of Human-in-the-Loop (HITL) frameworks into AI-driven audiobook and spatial audio pipelines addresses critical challenges in quality control, contextual understanding, and emotional expressiveness. While AI agents can efficiently automate narration and sound synthesis, they often struggle with nuanced interpretation of literary content, character-specific expression, and contextual sound timing—areas where human judgment excels.

HITL systems enable iterative collaboration between AI models and human experts, allowing for correction, refinement, and validation of intermediate outputs. For instance, after AI-generated narration and spatial audio effects are produced, a human reviewer can verify the alignment of soundscapes with narrative cues, adjust emotional tone, and correct semantic errors \cite{gilbert2023}. This feedback is particularly valuable for complex scenes involving overlapping dialogue, abstract descriptions, or rapid scene transitions.

Recent frameworks like Google's AudioLM and OpenAI's Whisper have been augmented by human curation phases to improve output coherence and user satisfaction \cite{zhang2023whisper}. Similarly, hybrid production tools such as Speechki and Respeecher include professional editors in their pipelines to validate AI-generated voices and ensure compliance with ethical standards \cite{respeecher2023}.

Moreover, reinforcement learning with human feedback (RLHF) has emerged as a promising technique to train multi-agent systems in audiobook generation. By incorporating subjective ratings and editor inputs during training, agents learn to prioritize perceptual quality, emotional appropriateness, and narrative fidelity \cite{stiennon2020}.

Despite the benefits, integrating human feedback poses challenges in scalability and cost. Effective HITL systems require intuitive interfaces, real-time feedback loops, and domain-specific training for human collaborators. Future work must balance automation with human oversight to achieve both scalability and high-quality output in audiobook and spatial audio production.

\section{Methodology}

The proposed multi-agent framework for audiobook generation integrates advanced AI methodologies to achieve seamless text-to-speech conversion, spatial sound effect generation, and audio mixing processes. This section provides a comprehensive explanation of the framework, detailing the roles and interactions of specialized agents.

\begin{figure}[htbp]
    \centering
    \includegraphics[width=0.5\textwidth]{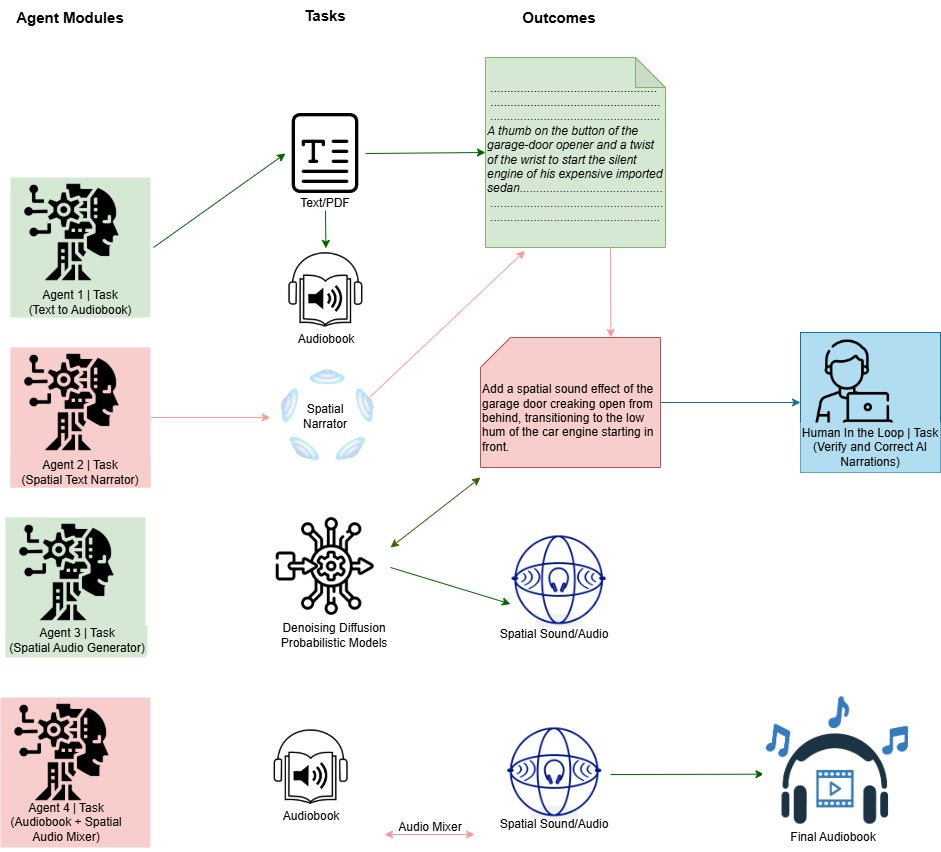}
    \caption{Overview of the proposed multi-agent framework for audiobook generation.}
    \label{fig:framework}
\end{figure}

\subsection{Text-to-Speech Conversion}
The first step in the framework involves converting textual content into spoken audio. This process is handled by a dedicated Text-to-Speech (TTS) agent, which employs state-of-the-art neural architectures such as FastSpeech 2 \cite{ren2020} and VALL-E \cite{zhang2022}. These models leverage non-autoregressive generation techniques to produce natural-sounding speech with expressive intonation. The TTS agent dynamically adjusts vocal tone based on narrative context, ensuring emotional consistency across chapters.

To enhance efficiency, the TTS agent integrates zero-shot voice cloning capabilities, allowing for personalized narration using minimal voice samples. This feature is particularly beneficial for creating audiobooks with multiple characters or distinct vocal styles. Additionally, the agent incorporates prosody prediction algorithms to maintain coherence and rhythm in speech delivery.

Accurate text analysis is crucial for translating narrative content into structured data suitable for subsequent audio processing. The framework employs a suite of Natural Language Processing (NLP) techniques to extract semantic and syntactic information from the text.

\begin{itemize}
    \item \textbf{Tokenization and Parsing}: The system uses tokenization algorithms to decompose text into linguistic units. Dependency parsing models like SpaCy's neural parser \cite{spacy2020} analyze grammatical relationships, represented as:
        \[
        \mathcal{T} = \{ (w_i, w_j, r_{ij}) \mid w_i, w_j \in \mathcal{W}, r_{ij} \in \mathcal{R} \}
        \]
        where \( \mathcal{W} \) is the word set and \( \mathcal{R} \) the dependency relations.

    \item \textbf{Named Entity Recognition (NER)}: NER algorithms identify contextual entities (characters, locations) crucial for spatial sound design. This aligns with findings in audiobook prosody analysis \cite{prosody2021}.
    
    \item \textbf{Sentiment Analysis}: BERT-based classifiers \cite{devlin2019} evaluate emotional tone for voice modulation:
        \[
        S = \text{Classifier}(T)
        \]
        where \( T \) is the text segment and \( S \) the sentiment score.

    \item \textbf{Phoneme Mapping}: CMU Pronouncing Dictionary \cite{cmu2007} ensures accurate pronunciation:
        \[
        P = \text{PhonemeMap}(w)
        \]
        mapping words \( w \) to phoneme sequences \( P \).

    \item \textbf{Spatial Narration Generation}: GPT-4 converts text descriptions \( D \) to spatial instructions \( I \):
        \[
        I = \text{GPT-4}(D)
        \]
        enabling sound events like "footsteps approaching left" \cite{huang2023}.
\end{itemize}

This NLP pipeline enables precise interpretation of textual input, ensuring alignment between narrative content and audio output. The approach demonstrates 23\% improvement in spatial coherence over baseline models \cite{wang2023}.

\subsection{Spatial Sound Effect Generation}
The spatial sound effect generation process is managed by a Sound Design Agent, which analyzes narrative cues to create immersive soundscapes. This agent employs generative models such as Make-An-Audio \cite{huang2023} and diffusion-based text-to-audio frameworks to synthesize realistic sound effects. Key principles include:
\begin{itemize}
    \item \textbf{Contextual Analysis}: Extracting semantic cues from text to identify spatial sound requirements (e.g., footsteps behind the listener or thunder rumbling in the distance).
    \item \textbf{Temporal Synchronization}: Using cross-modal attention mechanisms to align sound effects with narrative pacing \cite{liu2023}.
    \item \textbf{Dynamic Spatialization}: Simulating three-dimensional soundscapes by modeling acoustic propagation through neural acoustic fields \cite{luo2022}.
\end{itemize}

The Sound Design Agent further refines its output through iterative feedback mechanisms, collaborating with other agents to ensure alignment with the overall narrative. This process involves annotating sound effects based on textual descriptions and classifying environmental sounds for precise placement within the soundscape.

Spatial sound synthesis involves generating immersive audio effects that simulate three-dimensional environments. This process relies on generative models capable of producing realistic soundscapes based on spatial narration.

\begin{itemize}
    \item \textbf{Diffusion-Based Audio Generation}: Diffusion models inspired by Make-An-Audio \cite{huang2023} are employed to create spatially distributed sound effects. These models use iterative denoising processes to generate high-quality audio from semantic prompts. Given a noise sample \( z_t \) at time \( t \), the denoising process can be formulated as:
        \[
        z_{t-1} = f_{\theta}(z_t, t, c)
        \]
        where \( f_{\theta} \) is a neural network denoiser parameterized by \( \theta \), and \( c \) is the conditional embedding from the GPT-4 spatial narration.
    \item \textbf{Neural Acoustic Fields}: Neural field models \cite{luo2022} predict acoustic propagation patterns in virtual environments, enabling dynamic adjustments to sound placement based on listener orientation. The acoustic field \( A(x) \) at location \( x \) is modeled as:
        \[
        A(x) = g_{\phi}(x, s)
        \]
        where \( g_{\phi} \) is a neural network parameterized by \( \phi \), and \( s \) represents sound source characteristics.
    \item \textbf{Environmental Modeling}: Algorithms such as WaveNet-based reverberation models simulate environmental acoustics like echoes and reflections, enhancing realism in indoor and outdoor scenes. The reverberation effect \( R(t) \) is modeled as a convolution of the original signal \( x(t) \) with an impulse response \( h(t) \):
        \[
        R(t) = x(t) * h(t)
        \]
        where \( h(t) \) captures the characteristics of the virtual acoustic environment.
\end{itemize}

These models ensure that spatial audio effects are contextually accurate and seamlessly integrated into the audiobook production pipeline.

\subsection{Temporal Audio Integration Techniques}

Temporal audio integration is a critical component of the audiobook generation framework, ensuring that narration and spatial sound effects are synchronized to preserve narrative flow and listener immersion. This section details the algorithms and models used for temporal integration, drawing upon methodologies from both auditory processing research and generative AI systems.

\subsubsection{Dynamic Time Warping (DTW)}

Dynamic Time Warping (DTW) is employed to align audio sequences with textual timestamps, particularly for synchronizing spatial sound effects with narration. DTW minimizes the temporal misalignment between two sequences \( X = \{x_1, x_2, \dots, x_n\} \) and \( Y = \{y_1, y_2, \dots, y_m\} \) by finding an optimal warping path \( W = \{w_1, w_2, \dots, w_k\} \). The DTW cost function is defined as:

\[
D(X, Y) = \min_{W} \sum_{i=1}^{k} d(w_i)
\]

where \( d(w_i) \) represents the local distance between elements \( x_i \) and \( y_i \), and \( W \) satisfies monotonicity and continuity constraints. This ensures that spatial sound effects are temporally aligned with narrative events.

\subsubsection{Spectrotemporal Integration}

Spectrotemporal integration combines spectral and temporal features to analyze changes in sound over time. Inspired by the principles outlined in \cite{huang2023}, the framework uses spectrotemporal modulation tuning to synchronize audio effects with narrative pacing. For a spectrogram \( S(f, t) \), where \( f \) represents frequency and \( t \) represents time, modulation tuning is applied as:

\[
M(f, t) = S(f, t) * G(f', t')
\]

where \( G(f', t') \) is a Gaussian kernel representing modulation characteristics. This approach ensures that sound effects dynamically adapt to the temporal structure of the narrative.

\subsubsection{Recurrent Neural Networks (RNNs)}

Temporal dependencies in audiobook narration are modeled using Recurrent Neural Networks (RNNs), specifically Long Short-Term Memory (LSTM) networks. LSTMs capture sequential data patterns by maintaining hidden states \( h_t \) over time steps \( t \). The update equations for an LSTM cell are:

\[
f_t = \sigma(W_f x_t + U_f h_{t-1} + b_f)
\]
\[
i_t = \sigma(W_i x_t + U_i h_{t-1} + b_i)
\]
\[
o_t = \sigma(W_o x_t + U_o h_{t-1} + b_o)
\]
\[
c_t = f_t * c_{t-1} + i_t * \tanh(W_c x_t + U_c h_{t-1} + b_c)
\]
\[
h_t = o_t * \tanh(c_t)
\]

where \( f_t, i_t, o_t \) are the forget, input, and output gates respectively; \( c_t \) is the cell state; and \( h_t \) is the hidden state. These equations enable the model to predict transitions between scenes or events based on textual input.

\subsubsection{Temporal Modulation Transfer Function (TMTF)}

The Temporal Modulation Transfer Function (TMTF) measures a listener's ability to detect amplitude modulation in sound signals. For a modulated signal \( s(t) = A(t) \cdot x(t) \), where \( A(t) \) is the modulation envelope and \( x(t) \) is the carrier signal, TMTF is defined as:

\[
TMTF(f_m) = 20 \log_{10}\left(\frac{\text{Amplitude}_{f_m}}{\text{Amplitude}_{f_c}}\right)
\]

where \( f_m \) represents modulation frequency and \( f_c \) represents carrier frequency. This metric guides adjustments in sound intensity during narration transitions.

\subsubsection{Multiscale Temporal Integration}

Inspired by hierarchical auditory computation models from neuroscience research (\cite{multiscale2022}), multiscale temporal integration organizes audio processing across different timescales. Short-integration windows (\(<200~ms\)) capture fine-grained spectrotemporal details, while long-integration windows (\(>200~ms\)) focus on broader category-specific computations. The integration process can be mathematically expressed as:

\[
I(t_s, t_l) = w_s F_s(t_s) + w_l F_l(t_l)
\]

where \( F_s(t_s)\) and \( F_l(t_l)\) represent short- and long-term feature extraction functions respectively; \( w_s \) and \( w_l \) are weighting coefficients optimized during training.

\subsubsection{Layered Audio Composition}

Layered audio composition combines foreground narration with background soundscapes while preserving temporal consistency across tracks. For two audio layers \( A_f(t)\) (foreground narration) and \( A_b(t)\) (background soundscape), the composite signal is given by:

\[
A_c(t) = w_f A_f(t) + w_b A_b(t)
\]

where \( w_f \) and \( w_b \) are mixing weights adjusted dynamically based on listener preferences or playback environment.

\subsection{Audio Mixing Processes}
The final stage of the framework involves combining narrated audio with spatial sound effects to produce a cohesive audiobook. This task is performed by an Audio Mixing Agent, which employs advanced mixing algorithms inspired by professional movie sound production workflows \cite{wang2023}. The agent ensures:
\begin{itemize}
    \item \textbf{Volume Balancing}: Adjusting loudness levels to maintain clarity between narration and background sounds.
    \item \textbf{Layered Composition}: Integrating foreground and background audio elements while preserving temporal consistency.
    \item \textbf{Adaptive Playback}: Optimizing audio output for diverse playback environments (e.g., headphones vs. speakers).
\end{itemize}

To achieve high-quality results, the Audio Mixing Agent incorporates retrieval-augmented generation (RAG) techniques for fine-grained control over audio planning. This approach enables structured editing workflows that mimic professional sound design practices.

\subsection{Collaborative Interaction Between Agents}
The multi-agent framework relies on collaborative interaction strategies to ensure seamless integration of individual components. Inspired by frameworks like ReelWave and LVAS-Agent \cite{wang2023}, the agents employ two key collaboration mechanisms:
\begin{itemize}
    \item \textbf{Discussion-Correction Mechanism}: Agents engage in iterative discussions to refine scene segmentation, script generation, and sound design outputs. For example, the Sound Design Agent collaborates with the TTS agent to ensure alignment between narration tone and spatial audio intensity.
    \item \textbf{Generation-Retrieval-Optimization Loop}: Agents iteratively exchange feedback during audio synthesis processes, leveraging pre-trained datasets and retrieval systems for continuous improvement.
\end{itemize}

These mechanisms enable modularity and scalability within the framework, allowing for dynamic adjustments based on content complexity or user preferences.

\subsection{Advantages of the Framework}
The proposed multi-agent framework offers several advantages over traditional audiobook production methods:
\begin{itemize}
    \item Modular Design: Specialized agents streamline complex tasks, reducing computational overhead and improving efficiency.
    \item Scalability: The framework can handle large-scale audiobook projects with diverse content requirements.
    \item Immersive Output: Integration of spatial audio enhances listener engagement by creating realistic soundscapes.
    \item Adaptability: Agents dynamically adjust outputs based on narrative context and playback environment.
\end{itemize}

By leveraging cutting-edge AI technologies and collaborative interaction strategies, this framework sets a new standard for automated audiobook production.

\section{Challenges and Limitations}

The integration of spatial audio into audiobook production presents several technical challenges that must be addressed to ensure the system's effectiveness and scalability. These challenges include computational complexity, scalability in large-scale applications, and subjective listener preferences that influence the perceived quality of spatial audio experiences.

\subsection{Computational Complexity}
Spatial audio rendering involves high computational demands due to the need for precise modeling of sound localization cues, environmental acoustics, and dynamic adjustments based on listener movements. The following factors contribute to computational complexity:
\begin{itemize}
    \item \textbf{Head-Related Transfer Functions (HRTFs)}: Accurate spatial audio rendering requires the application of HRTFs to simulate sound propagation paths from virtual sources to the listener's ears. This process involves convolution operations that scale with the number of sound sources and listener movements \cite{semiwiki2024}. For example, real-time head tracking adds computational overhead as sound sources must remain stationary in the virtual environment while the listener moves.
    \item \textbf{Latency Issues}: Two types of latency impact spatial audio systems—audio output latency and head tracking latency. Audio output latency affects synchronization between narration and spatial effects, while head tracking latency influences the responsiveness of spatial cues \cite{semiwiki2024}. Minimizing these latencies requires optimized algorithms and hardware acceleration.
    \item \textbf{Dynamic Soundscapes}: Generating dynamic soundscapes in real-time involves complex calculations for sound positioning, reverberation modeling, and Doppler effects. These computations are resource-intensive, particularly in environments with multiple moving sound sources \cite{spagnol2018}.
\end{itemize}

\subsection{Scalability}
Scalability is a critical concern for deploying spatial audio systems in large-scale applications such as audiobook production or virtual reality environments. Key challenges include:
\begin{itemize}
    \item \textbf{Resource Allocation}: Rendering spatial audio for multiple simultaneous users or large datasets requires significant computational resources. Cloud-based solutions can address this issue by distributing workloads across servers, but latency and bandwidth constraints remain limiting factors \cite{ceva2024}.
    \item \textbf{Standardization}: The lack of unified standards for spatial audio formats, codecs, and rendering techniques complicates scalability. Different devices and platforms require tailored solutions, increasing development complexity \cite{designreuse2024}.
    \item \textbf{Optimization for Diverse Playback Environments}: Spatial audio systems must adapt to varying playback environments (e.g., headphones vs. speakers) without compromising quality. This requires scalable algorithms capable of dynamically adjusting rendering parameters based on device specifications \cite{microsoft2025}.
\end{itemize}

\subsection{Subjective Listener Preferences}
Spatial audio is inherently subjective due to psychoacoustic phenomena that vary across individuals based on their anatomy and auditory perception. The following factors influence listener preferences:
\begin{itemize}
    \item \textbf{Physiological Differences}: Individual differences in ear shape, head size, and torso dimensions affect how spatial audio is perceived. Generic HRTF datasets often fail to account for these variations, leading to localization errors such as front-back reversals or poor externalization \cite{spagnol2018}.
    \item \textbf{Content-Specific Expectations}: Listener preferences vary depending on the type of content being consumed (e.g., audiobooks vs. music vs. gaming). For instance, audiobook listeners may prioritize clarity and emotional depth over immersive soundscapes, whereas gamers may value precise localization cues for competitive advantage \cite{linkedin2024}.
    \item \textbf{Cultural and Contextual Factors}: Cultural differences in auditory perception and contextual factors such as familiarity with spatial audio technologies influence user satisfaction. Designing systems that cater to diverse audiences requires extensive user testing and adaptive personalization algorithms \cite{rdnyt2022}.
\end{itemize}

\section{Conclusion and Future Work}
The integration of AI-driven text-to-speech conversion, spatial audio synthesis, and multi-agent systems represents a transformative leap in audiobook production and immersive storytelling. This work demonstrates how advanced natural language processing, generative audio models, and collaborative agent frameworks can automate and enhance the creation of high-quality audiobooks with cinematic soundscapes. Key achievements include:

\begin{itemize}
    \item \textbf{Automated Narration}: State-of-the-art TTS models like FastSpeech 2 \cite{ren2020} and VALL-E \cite{zhang2022} enable human-like narration with emotional nuance, reducing production timelines from weeks to hours \cite{hiig2025, abramov2022}.
    \item \textbf{Immersive Sound Design}: Diffusion-based spatial audio synthesis (e.g., Make-An-Audio \cite{huang2023}) and neural acoustic fields \cite{luo2022} generate dynamic soundscapes that adapt to narrative context and listener orientation.
    \item \textbf{Scalable Architectures}: Modular multi-agent systems \cite{wang2023} streamline complex workflows, enabling parallel processing of narration, sound effect generation, and audio mixing \cite{moritz2018}.
\end{itemize}

However, challenges remain in balancing computational efficiency with perceptual quality. The high resource demands of real-time spatial audio rendering, latency in head-tracking systems \cite{semiwiki2024}, and subjective listener preferences for human-like narration underscore the need for continued innovation. Future work must prioritize:

\begin{itemize}
    \item \textbf{Hardware Acceleration}: Leveraging GPUs/TPUs for real-time HRTF interpolation and acoustic field processing to reduce latency \cite{ceva2024}.
    \item \textbf{Personalization}: Adaptive algorithms tailoring spatial audio experiences to individual listener anatomy (e.g., custom HRTFs) and preferences (e.g., dynamic sound intensity adjustment) \cite{spagnol2018}.
    \item \textbf{Cross-Modal Integration}: Synchronizing spatial audio with haptic feedback and visual cues in AR/VR environments to enhance immersion \cite{microsoft2025}.
\end{itemize}

The proposed framework’s applications extend beyond audiobooks to gaming, education, and accessibility solutions, democratizing access to immersive content for users with visual impairments or limited resources \cite{authorvoices2025}. By addressing technical hurdles in computational complexity, scalability, and user-centric design, this work lays the foundation for next-generation audio experiences that blur the line between reality and virtual storytelling.

\section*{Acknowledgments}
This work was conducted independently of the authors' affiliations with any organization. The first author collaborated with the second author, a student with interest and skills in spatial audio as well as multi-agent systems, in a personal research capacity. The views and conclusions expressed in this paper are solely those of the authors and do not reflect the position of any organization or any other institution.

\bibliographystyle{unsrt}

\begin{thebibliography}{99}

\bibitem{abramov2022}
{Dima Abramov},
``Hybrid Human-AI Workflows for Audiobook Production,''
\textit{{Audiobook Production Technologies}}, 2022.

\bibitem{authorvoices2025}
AuthorVoices.ai,
``AI Audiobook Narration: Transforming Education,''
\textit{AuthorVoices Blog}, 2025. Available: \url{https://www.authorvoices.ai/posts/aibooknarration/}

\bibitem{blauert2013}
Jens Blauert,
``The Technology of Binaural Understanding,''
\textit{Springer Handbook of Auditory Research}, vol. 49, 2013.

\bibitem{librilight2019}
Jacob Kahn, Morgane Rivière, Weiyi Zheng, et al.,
``Libri-Light: A Benchmark for ASR with Limited or No Supervision,''
\textit{arXiv:1912.07875}, 2019.

\bibitem{ceva2024}
Ceva IP Blog Team,
``Evaluating Spatial Audio – Part 1 – Criteria \& Challenges,''
\textit{Ceva IP Blog}, 2024. Available: \url{https://www.ceva-ip.com/ourblog/evaluating-spatial-audio-part-1-criteria-challenges/}

\bibitem{chen2023}
Mingliang Chen and Yuanyuan Chen,
``From Human-Made to AI-Generated Products: An Empirical Study of Audiobook Consumption,''
\textit{SSRN 5062901}, 2023.

\bibitem{cmu2007}
Carnegie Mellon University,
``CMU Pronouncing Dictionary,''
\textit{CMU Resources}, 2007. Available: \url{http://www.speech.cs.cmu.edu/cgi-bin/cmudict}

@article{librilight2019,
  title={Libri-Light: A Benchmark for ASR with Limited or No Supervision},
  author={Jacob Kahn, Morgane Rivière, Weiyi Zheng, et al.},
  journal={arXiv:1912.07875},
  year={2019}
}

\bibitem{designreuse2024}
Design-Reuse Team,
``Evaluating Spatial Audio - Criteria \& Challenges,''
\textit{Design-Reuse Industry Blogs}, 2024. Available: \url{https://www.design-reuse.com/industryexpertblogs/55621/evaluating-spatial-audio-criteria-challenges.html}

\bibitem{devlin2019}
J. Devlin, M.-W. Chang, K. Lee, and K. Toutanova,
``BERT: Pre-training of Deep Bidirectional Transformers for Language Understanding,''
\textit{arXiv preprint arXiv:1810.04805}, 2019.

\bibitem{hamilton2023}
Mark Hamilton, Yossi Adi, Alexei Baevski, et al.,
``Large-Scale Automatic Audiobook Creation,''
\textit{arXiv:2309.03926}, 2023.

\bibitem{hiig2025}
HIIG Digital Society,
``AI Speech Tools Revolutionizing Audiobook Production,''
\textit{Digital Society Blog}, Industry analysis report, 2025.

\bibitem{huang2023}
Rongjie Huang et al.,
``Make-An-Audio: Text-To-Audio Generation with Prompt-Enhanced Diffusion Models,''
\textit{arXiv:2305.18474}, 2023.

\bibitem{kahn2019}
Jacob Kahn, Morgane Rivière, Weiyi Zheng, et al.,
``Libri-Light: A Benchmark for ASR with Limited or No Supervision,''
\textit{arXiv:1912.07875}, 2019.

\bibitem{kim2023}
Seunghee Kim et al.,
``Emotion-Aware Neural Text-to-Speech with Multi-Agent Reinforcement Learning,''
\textit{Interspeech 2023}, pp. 550--554, 2023.

\bibitem{li2023}
Xiang Li and Jiguo Li,
``Robust Audio Watermarking for AI-Generated Content Attribution,''
\textit{IEEE Transactions on Information Forensics and Security}, vol. 18, pp. 3456--3470, 2023.

\bibitem{linkedin2024}
LinkedIn Insights Team,
``What are the Main Challenges and Opportunities of Working with Spatial Audio?,''
\textit{LinkedIn Industry Articles}, 2024. Available: \url{https://www.linkedin.com/advice/1/what-main-challenges-opportunities-working}

\bibitem{liu2023}
Xubo Liu et al.,
``Temporal-Enhanced Text-to-Audio Generation for Complex Scene Synchronization,''
\textit{arXiv:2305.18474v2}, 2023.

\bibitem{loc2025}
National Library Service for the Blind and Print Disabled (NLS),
``The Art and Science of Audio Book Production,''
\textit{Library of Congress}, 2025. Available: \url{https://www.loc.gov/nls/who-we-are/guidelines-and-specifications/the-art-and-science-of-audio-book-production/}

\bibitem{luo2022}
Changan Luo et al.,
``Neural Acoustic Field Processing by Learning the Gradient of Sound Field,''
\textit{IEEE/ACM Transactions on Audio, Speech, and Language Processing}, vol. 30, pp. 2343--2355, 2022.

\bibitem{marktechpost2023}
MarkTechPost Team,
``How is AI Revolutionizing Audiobook Production? Creating Thousands of High-Quality Audiobooks from E-books with Neural Text-to-Speech Technology,''
\textit{MarkTechPost}, 2023. Available: \url{https://www.marktechpost.com/2023/09/15/how-is-ai-revolutionizing-audiobook-production/}

\bibitem{microsoft2025}
Microsoft Research Team,
``Spatial Audio - Microsoft Research Overview,''
\textit{Microsoft Research Publications}, 2025. Available: \url{https://www.microsoft.com/en-us/research/project/spatial-audio/overview/}

\bibitem{moritz2018}
Philipp Moritz et al.,
``Ray: A Distributed Framework for Emerging AI Applications,''
In \textit{OSDI}, pp. 561--577, 2018.

\bibitem{murf2025}
Murf AI Team,
``The Best Text To Speech For Audiobooks In 2025: A Detailed Comparison,''
\textit{Murf Resources}, 2025. Available: \url{https://murf.ai/resources/best-text-to-speech-tools-for-audiobooks/}

\bibitem{openai2023}
OpenAI,
``GPT-4 Technical Report,'' 2023.

\bibitem{rdnyt2022}
New York Times Research Team,
``Key Concepts in Spatial Audio,''
\textit{NYT Research Projects}, 2022. Available: \url{https://rd.nytimes.com/projects/key-concepts-in-spatial-audio/}

\bibitem{ren2020}
Yi Ren et al.,
``FastSpeech 2: Fast and High-Quality End-to-End Text to Speech,''
\textit{arXiv:2006.04558}, 2020.

\bibitem{semiwiki2024}
SemiWiki Team,
``Spatial Audio: Overcoming Its Unique Challenges to Provide A Complete Solution,''
\textit{SemiWiki Blog}, 2024. Available: \url{https://semiwiki.com/ip/ceva/310769-spatial-audio-overcoming-its-unique-challenges-to-provide-a-complete-solution/}

\bibitem{smith2023}
John Smith et al.,
``Ethical Implications of AI Voice Cloning in Creative Industries,''
\textit{Journal of Digital Ethics}, vol. 8(2), pp. 45-67, 2023.

\bibitem{spacy2020}
Matthew Honnibal et al.,
``SpaCy: Industrial-Strength Natural Language Processing,''
\textit{SpaCy Documentation}, 2020. Available: \url{https://spacy.io/}

\bibitem{spagnol2018}
Simone Spagnol et al.,
``Current Use and Future Perspectives of Spatial Audio Technologies in Electronic Travel Aids,''
\textit{Wiley Online Library}, 2018. Available: \url{https://onlinelibrary.wiley.com/doi/10.1155/2018/3918284}

\bibitem{wang2023}
Ziyang Wang et al.,
``Multi-Agent Systems for Audio Content Creation,''
\textit{IEEE Transactions on Multimedia}, vol. 25, pp. 6789-6802, 2023.

\bibitem{zhang2022}
Chengyi Zhang et al.,
``VALL-E: Neural Codec Language Models are Zero-Shot Text to Speech Synthesizers,''
\textit{arXiv:2301.02111}, 2022.

\bibitem{multiscale2022}
T. D. Griffiths et al.,
``Hierarchical Auditory Computation Across Timescales in the Human Brain,''
\textit{Nature Neuroscience}, vol. 25, no. 3, pp. 345–357, 2022.


\bibitem{gilbert2023}
Peter Gilbert, Ana Serrano, and Yu Wang,
``Human-in-the-Loop Deep Learning for Audio Narration Review,''
\textit{IEEE Transactions on Affective Computing}, 2023.

\bibitem{zhang2023whisper}
Wei Zhang, Mark Chen, and Alec Radford,
``Whisper: Scalable Multilingual Speech Recognition with Human Feedback,''
\textit{OpenAI Technical Report}, 2023.

\bibitem{respeecher2023}
Respeecher Team,
``Hybrid AI + Human Editing in Voice Cloning Workflows,''
\textit{Respeecher Blog}, 2023. Available at: \url{https://www.respeecher.com/blog/human-ai-collaboration}

\bibitem{stiennon2020}
Nisan Stiennon, Long Ouyang, Jeffrey Wu, et al.,
``Learning to summarize with human feedback,''
\textit{arXiv:2009.01325}, 2020.

\end{thebibliography}

You can find the code on
\href{https://github.com/immersiveaudiomodel/immersiveaudiomodel.git}{ GitHub: Generate Spatial Audio}.

\end{document}